\documentclass{article}
\begin{document}
\date{October 15, 2002}
\title{ 
Searches for Leptoquarks with the D\O~Detector at the TeVatron}
\author{
T.~Christiansen 
\\
\em LMU M\"unchen, Am Coulombwall 1, D-85748 Garching, Germany
}
\maketitle
\baselineskip=11.6pt
\begin{center}
\large
Poster Presentation \& Short Communication in Plenary Session of \\
\Large Frontier Science, Frascati, Italy, Oct.~6-11, 2002
\end{center}
\baselineskip=14pt
%
The existence of leptoquarks (LQ), color-triplets of bosons with lepton 
and quark quantum numbers,
is predicted by different theories beyond the Standard 
Model. 
This article summarizes the D\O~Run-I searches for the three different 
LQ generations in $p\bar{p}$ collisions at
$\sqrt{s} = 1.8$ TeV and presents ongoing studies at Run II.

\begin{figure}[h]
 \vspace{-.6cm}
%
%
\includegraphics{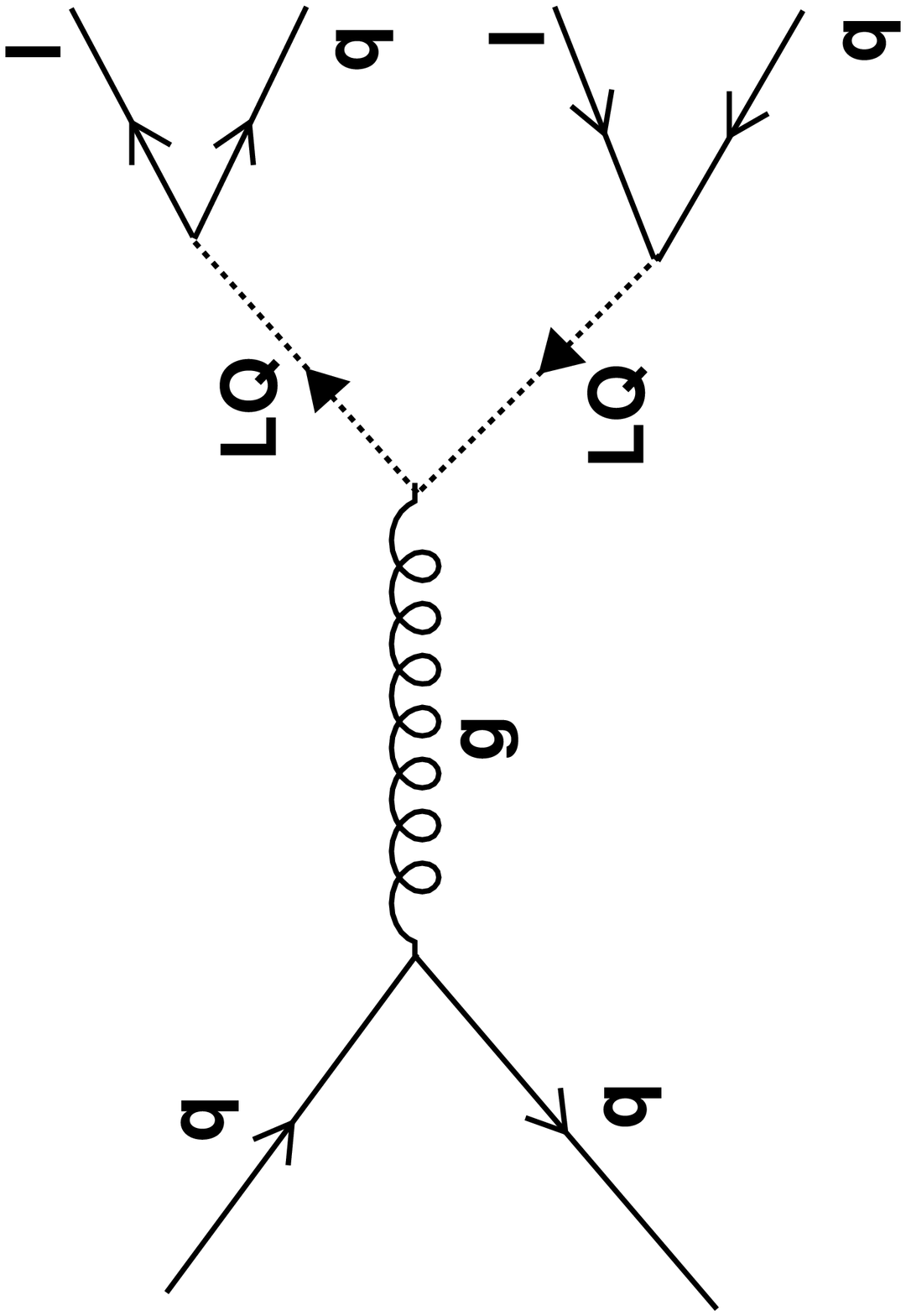}
\includegraphics{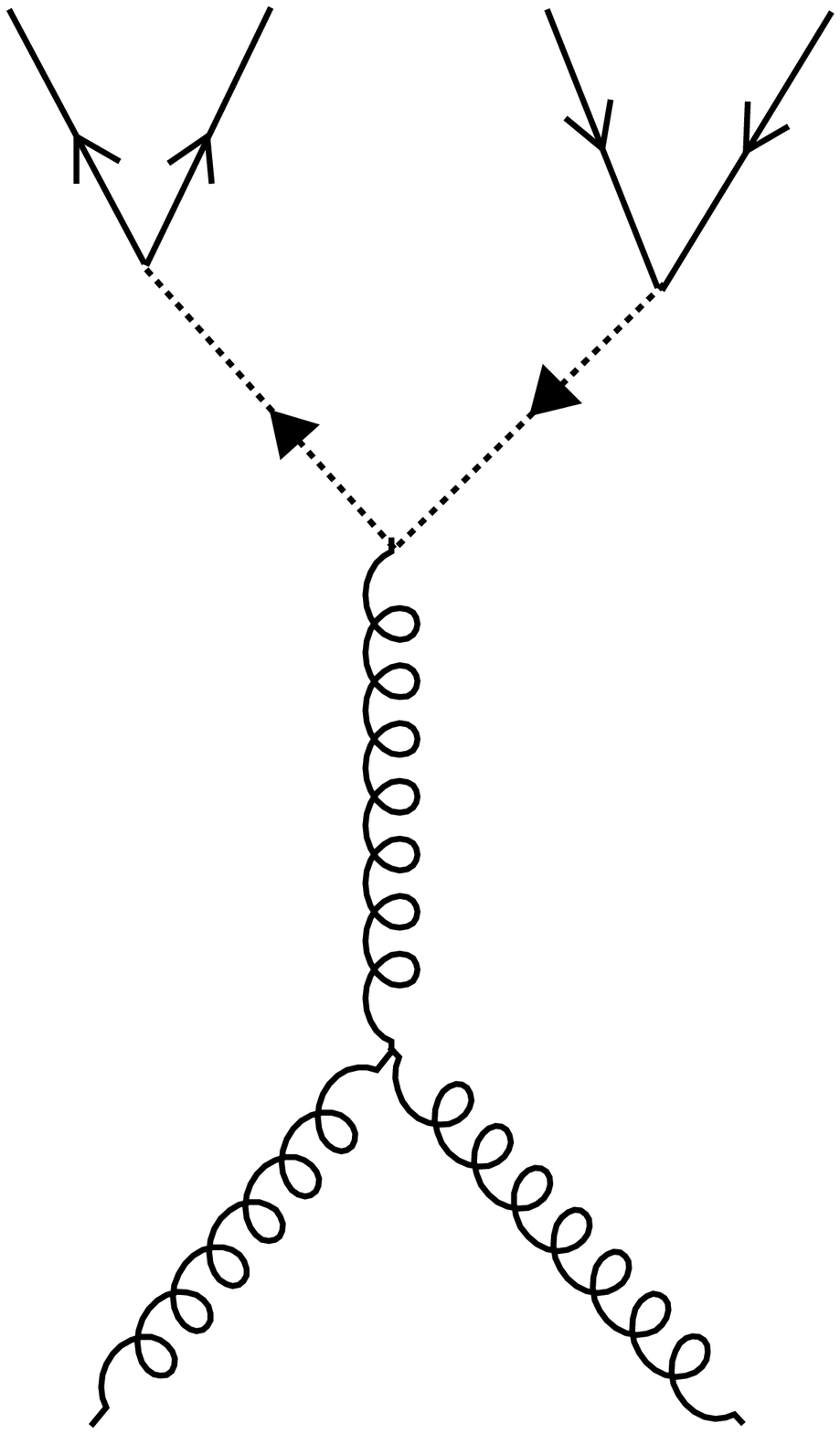}
\includegraphics{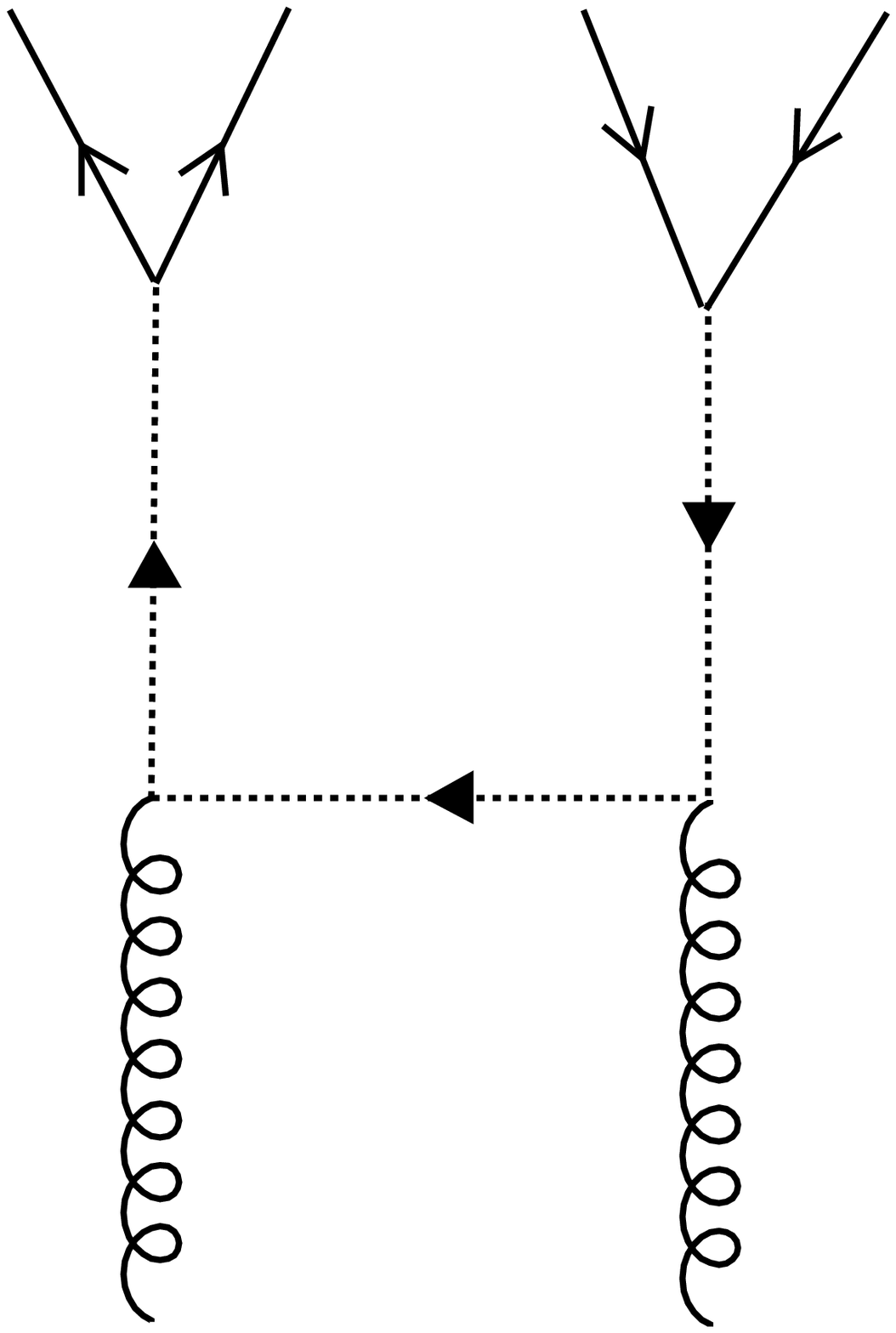}
\includegraphics{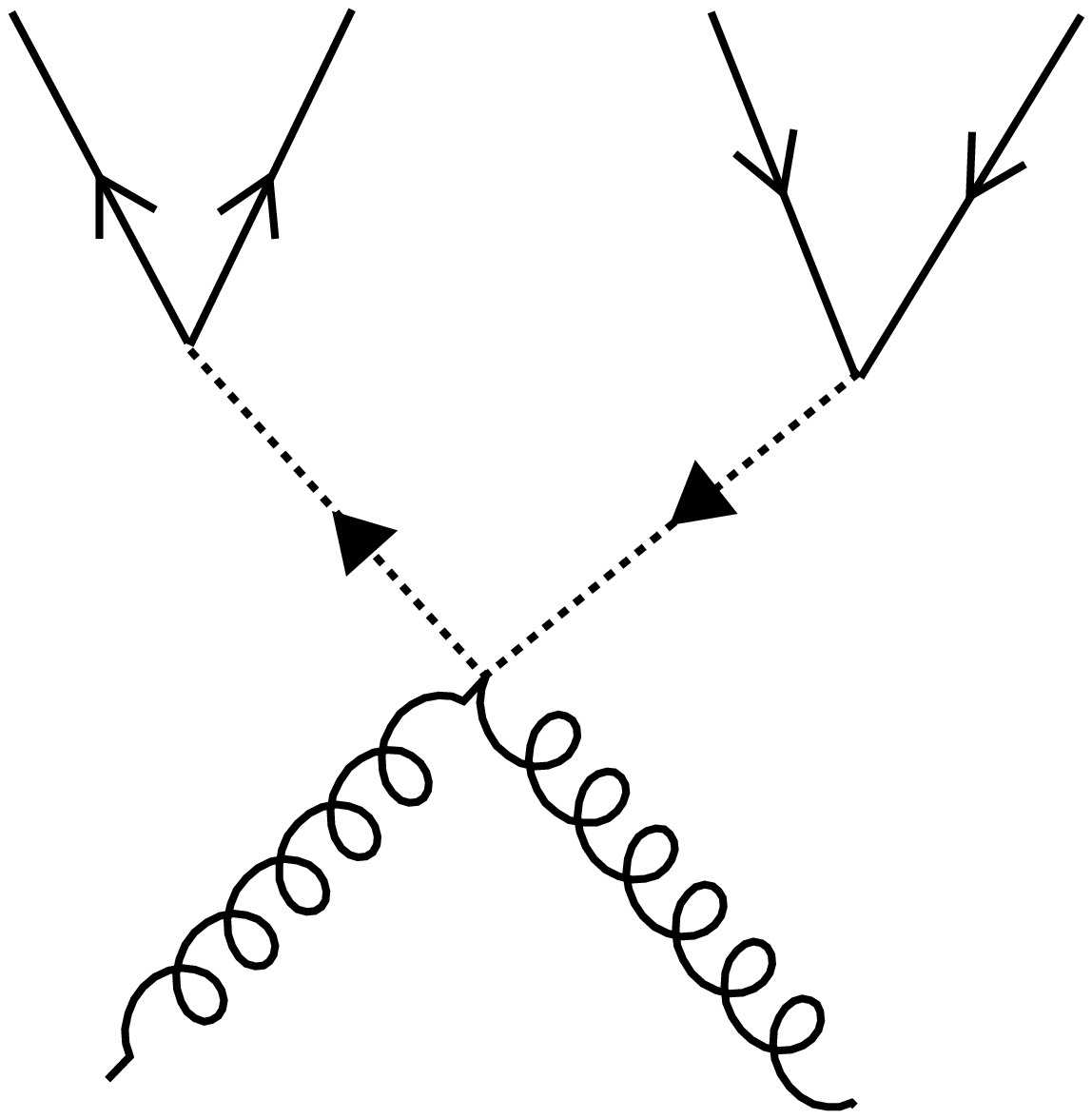}
 \vspace{2.1cm}
 \caption{
      Leading-order diagrams for leptoquark pair production.
    \label{feynman} }
\end{figure}

%
Limits on proton decays, on lepton
flavor violation and on flavor-changing neutral currents lead to 
the assumption
that there would be three different generations of leptoquarks, each one of 
them only interacting within one lepton and quark family. 
Figure \ref{feynman}
shows Feynman graphs of the dominant production processes in $p\bar{p}$ 
collisions.
 \vspace{-0.3cm}

Using about 100 pb$^{-1}$ of $p\bar{p}$ collision data with a center-of-mass 
energy of $\sqrt{s}=1.8$ GeV recorded during Run I with the D\O~detector, 
no evidence for the existence of leptoquarks was found.
The results on the leptoquark searches were combined to 95\% confidence 
limits on the LQ-pair production cross section as a function of the mass
$m_{LQ}$ for the different models and leptoquark generations. 
The lower limits for masses of 1$^{\rm{st}}$-generation scalar leptoquarks are 
225 GeV, 204 GeV and 79 GeV, assuming a branching fraction to charged leptons
of $\beta=BF(LQ\rightarrow l^\pm q)$ = 1, 1/2 and 0, respectively.\cite{runI_first}
The corresponding
limits for the 2$^{\rm{nd}}$-generation scalar LQ masses are determined 
to 200 GeV, 180 GeV and 79 GeV.\cite{runI_second}
For 3$^{\rm{rd}}$-generation LQ masses comparable to $m_{top}$, the decay 
$LQ_3\rightarrow t+l$ is suppressed or even forbidden.
Studies of the $\nu b\bar{\nu} \bar{b}$
channel yield a lower mass limit of $94$ GeV for scalar 3$^{\rm{rd}}$-generation leptoquarks and $\beta=0$.\cite{runI_third}

\begin{figure}[h]
 \vspace{2.6cm}
%
%
\includegraphics{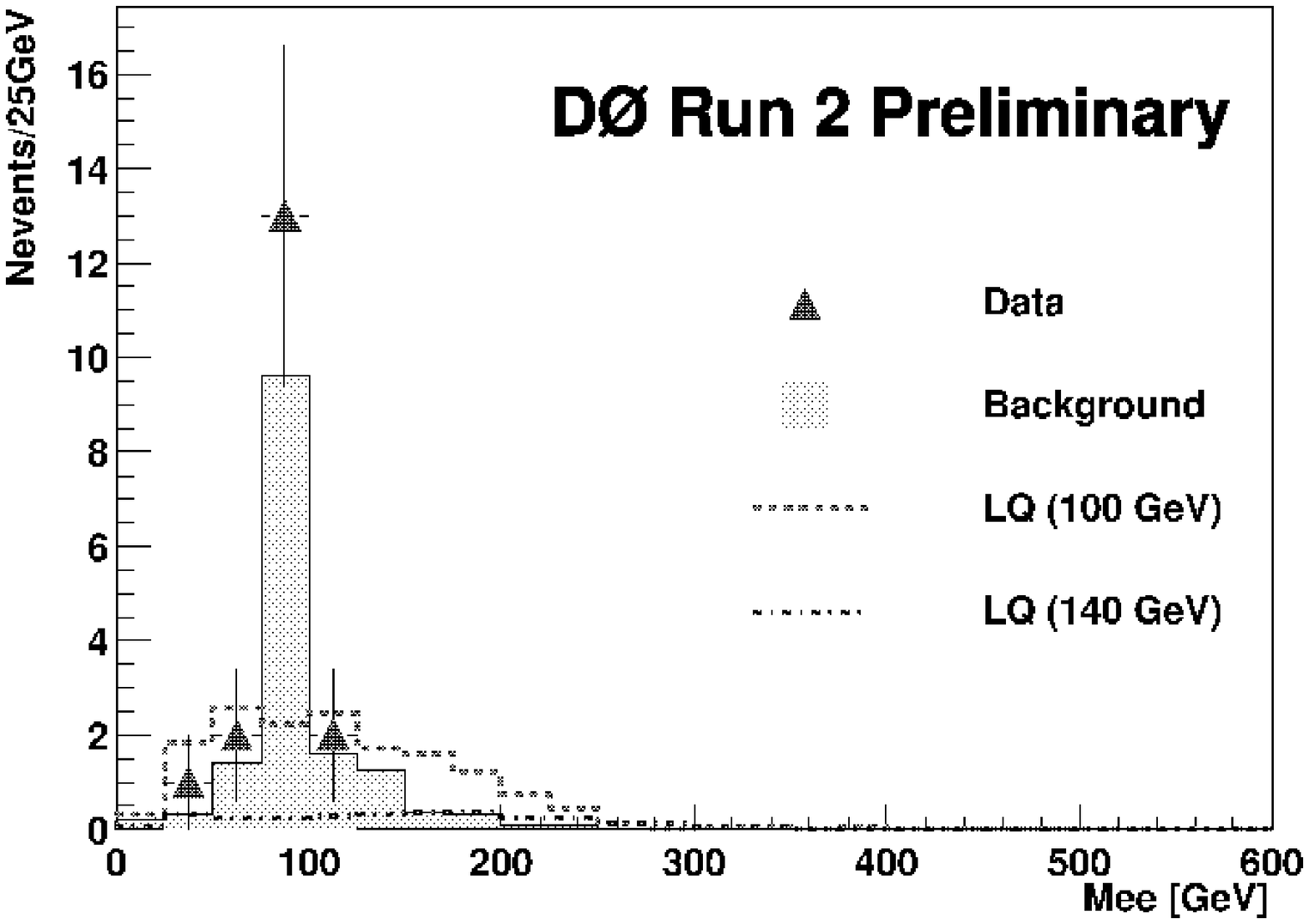}
\includegraphics{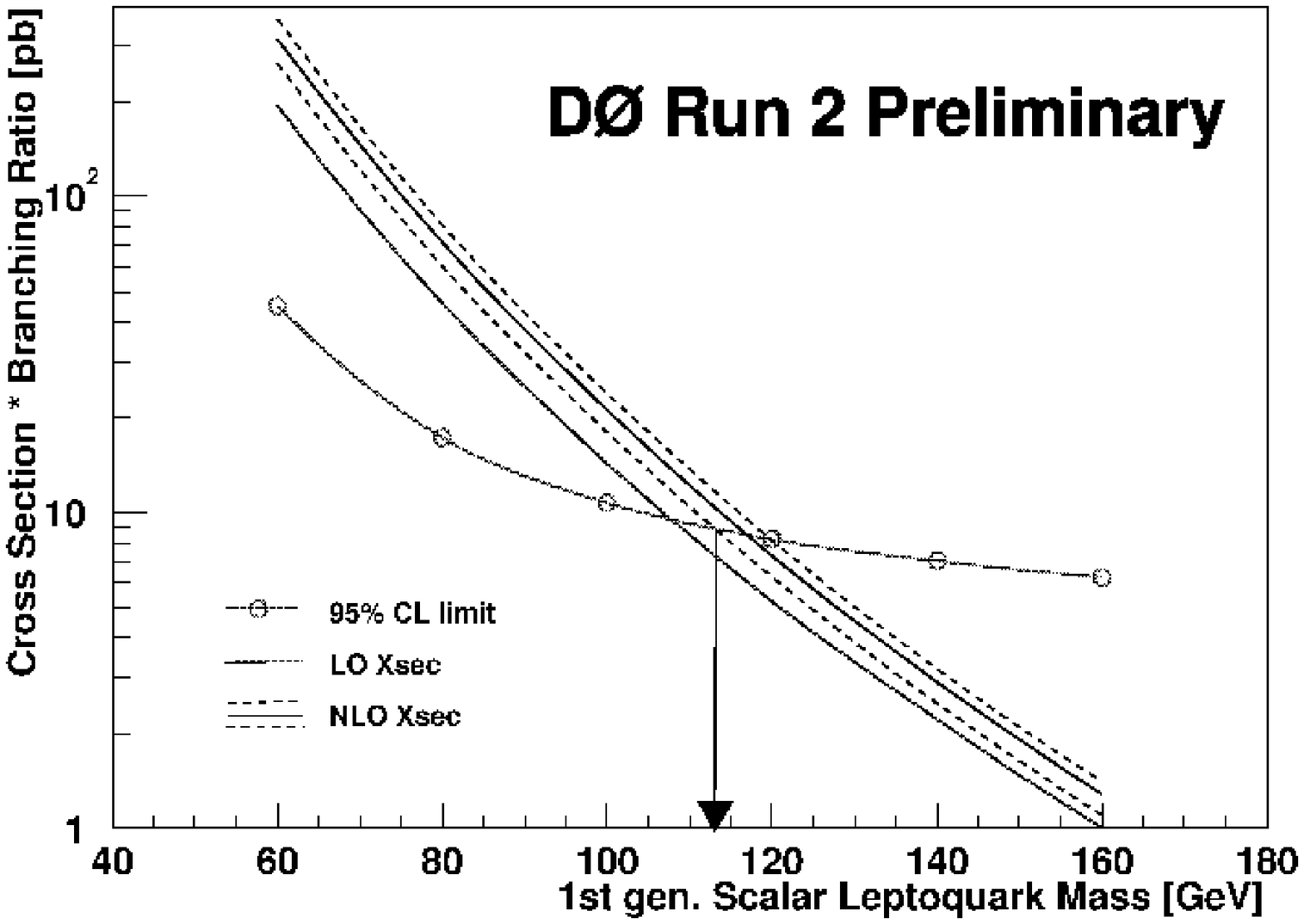}
 \vspace{0.3cm}
 \caption{
      Search for 1$^{\rm{st}}$-generation leptoquarks at Run II. Left: Di-electron 
      invariant mass of $eejj$ candidate events. Right: LQ-mass dependent limits on the cross section 
      for scalar LQ pairs 
      assuming 100\% branching fraction to charged leptons 
      $BF(LQ\rightarrow eq)=1$.\cite{RunII_first}
    \label{runII} }
\end{figure}

If both leptoquarks decay into an electron and a quark, 
two electrons and two jets can be reconstructed.
The left diagram in figure \ref{runII} shows the di-electron 
mass of $eejj$ Run-II candidates collected 
between November 2001 and May 2002 
with $\sqrt{s}\approx 2$ GeV. These 18
events, equivalent to an integrated luminosity of about $8.8$ pb$^{-1}$, are 
compatible with background ($15\pm5$ events expected) which is dominated by 
Drell-Yan $Z$ production.
Comparing the limits on the cross section for different masses $m_{LQ}$ to
NLO calculations for scalar leptoquarks, the lower limit on the mass of 
1$^{\rm{st}}$-generation leptoquarks is 113 GeV.\cite{RunII_first}
This is compatible with
earlier results from Run I and reflects the lower integrated luminosity.
\vspace{-0.3cm}

Taking advantage of the increased collision energy and
luminosity of the TeVatron for Run II, D\O~will be able to extend its
searches for leptoquarks to so far inaccessible leptoquark masses.
Especially for the second generation,
the search for leptoquark pair production
will greatly benefit from the upgraded muon spectrometer,
the central tracking system with the newly installed magnetic solenoid
and the new trigger system.

\end{document}